\documentclass[11pt,a4paper]{article}
\usepackage[]{amsmath,amssymb}
\usepackage{graphics,epsfig}
\begin{document}
\date{}
\title{Relative entropy and the Bekenstein bound}
\author{H. Casini\footnote{e-mail: casini@cab.cnea.gov.ar} \\
{\sl Centro At\'omico Bariloche,
8400-S.C. de Bariloche, R\'{\i}o Negro, Argentina}}
\maketitle

\begin{abstract}
Elaborating on a previous work by Marolf et al, we relate some exact results in quantum field theory and statistical mechanics to the Bekenstein universal bound on entropy. Specifically, we consider the relative entropy between the vacuum and another state, both reduced to a local region. We propose that, with the adequate interpretation, the positivity of the relative entropy in this case constitutes a well defined statement of the bound in flat space. We show that this version arises naturally from the original derivation of the bound from the generalized second law when quantum effects are taken into account. In this formulation the bound holds automatically, and in particular it does not suffer from the proliferation of the species problem. The results suggest that while the bound is relevant at the classical level, it does not introduce new physical constraints semiclassically. 
\end{abstract}

\section{Introduction}
 The Bekenstein bound is a proposal for a universal bound on the entropy of a region in flat space.
 If $S$ is the entropy, $E$ is the total energy, and $R$ is some typical size of the system, it writes
\begin{equation}
S \le \lambda E\, R \,,\label{uno}
\end{equation}
where $\lambda$ is a numerical constant of order one.

The bound was originally conceived through a thought experiment involving black hole thermodynamics and classical physics \cite{bekbb}. This is as follows. Consider an object much smaller than the black hole radius and drop it from a position outside the black hole. It will be swallowed and disappear behind the horizon carrying away its entropy. The generalized second law of thermodynamics implies that the lost entropy have to be compensated (or exceeded) by an increase in the black hole area \cite{gsl}. This is in turn governed by the Einstein equations for gravity, and depends on the energy absorbed by the black hole, leading to (\ref{uno}).  

This derivation of the bound from the generalized second law of thermodynamics  (GSL) was impugned \cite{crit,marosork} and defended \cite{def,bekdowe} in several occasions.    
However, despite this controversy around its association with the GSL, the bound itself is interesting in its own right. The statement of the inequality (\ref{uno}) does not mention anything outside the area of ordinary flat space-time physics.  After looking at it a bit closer one may convince oneself that what is behind protecting the bound are primarily the uncertainty relations (except for the proliferation of the species problem) \cite{uncertainty}. This suggests that a full proof could be attempted in quantum field theory (QFT).

However, the first and more difficult step in such direction is to properly define the quantities involved. In fact, the bound has been successfully tested in many different quantum systems, but the calculations have always involved at least some heuristic input.

The main obstacle in properly stating the bound is the same that impeded the early attempts to construct a relativistic quantum theory for single particle (rather than a QFT). Eq. (\ref{uno}) requires defining a system which is bounded in space. This traduces quantum mechanically in the necessity of counting with a coordinate operator $X$. However, in a relativistic theory this would imply the existence of a time operator, a notion which clashes with quantum mechanics \cite{toperator}. 

Alternatively, one can try to define the localization of the system by imposing that the expectation values for local operators (i.e. the energy momentum tensor) in the region outside the one where the system is located coincide with the corresponding vacuum expectation values \cite{pagedef}. The entropy would in that case be given by its microcanonical expression $S=\log N(E_0)$ where $N(E_0)$ is the dimension of the Hilbert space of states which fulfill these conditions, with energy below $E_0$. However, the set of states selected in this manner does not form a vector space, and the entropy cannot be computed \cite{haag}. Remarkably, the Reeh-Schlieder theorem implies that the vector space generated by this set of states by taking linear combinations span the whole set of states with $E \le E_0$ (including even very far located particles). The reason for   this fact is that the states "localized" in this way are not really completely vanishing outside $V$. They have small tails outside $V$ which can be exploited by adequately combining them (for a review on the problem of localization in QFT see \cite{localization}, for a different localization scheme see \cite{piazza}, and for a review on the difficulties in correctly defining the bound see \cite{boudefining}).

Therefore, we can compute the energy and the entropy for a global state, but in that case we do not know what to make of $R$ in (\ref{uno}). On the contrary, localizing the state in a QFT inevitably leads to pair creation, and the entropy and energy became ill defined. In a specific calculation this translates into divergences (think for example on the infinite entropy carried by the Unruh radiation in Rindler space). In order to eliminate them we can either impose a cutoff or ad hoc boundary conditions which eliminate the correlations of the localized state with the rest of the space. In both cases the result contains input from external elements and is ambiguous.  

A way out of this dilemma was discovered in \cite{maro} (see also \cite{walda}). There it was observed that a more reasonable measure for the entropy carried by the system in a region of the space is obtained by subtracting to the bare entropy of the local state the entropy corresponding to the vacuum fluctuations. This last one is present in absence of any energy, and is entirely due to the localization. The subtracted entropy turns out to be finite. 

A similar subtraction can be done to define a localized form of energy.  Using this, we show how to define a version of the Bekenstein bound which has sense in quantum field theory. This coincides with the positivity of the relative entropy between two local states. We also show that this version of the bound emerges from known proofs of the GSL in the semiclassical quasistatic regime.

The type of ideas involved in this work appears also in this context in the refs. \cite{marosork,sork1,sork2,page,turn}, and especially in \cite{maro}, which largely motivated this paper. We hope it could help to clarify some aspects of the discussion in these works.

\section{A localized entropy}

First we focus on the left hand side of the bound inequality (\ref{uno}). Let us take a spatial region $V$ lying on a Cauchy surface. In a quantum field theory there is an algebra of local operators ${\cal A} (V)$ associated to $V$ (or more precisely to its causal development). If $\rho$ is the global state of the system, its localization $\rho_V$ to the region $V$ is given by the restriction of $\rho$ to ${\cal A} (V)$. In the continuum theory, it is not possible to give an entropy for $\rho_V$ since in general ${\cal A} (V)$ is a von Neumann algebra of a type which does not admit a trace \cite{haag}. However, let us think in terms of a cutoff theory (i.e. with a lattice regularization), in such a way that the Hilbert space can be decomposed as a tensor product ${\cal H}={\cal H}_V\otimes {\cal H}_{-V}$ of Hilbert spaces, associated to $V$ and its complementary set $-V$ on a Cauchy surface. These later give a product representation of the commuting algebras ${\cal A} (V)$ and ${\cal A} (-V)$. Then we have
\begin{equation}
\rho_V=\textrm{tr}_{-V}\,\rho\,,
\end{equation}
where the trace is over ${\cal H}_{-V}$ (that is, over the degrees of freedom lying outside $V$). In the limit of vanishing distance cutoff this state has a divergent entropy \cite{bom}
\begin{equation}
S(\rho_V)=-\textrm{tr} \rho_V \log \rho_V\,.
\end{equation}
In order to see this, consider for example the global vacuum reduced to the Rindler wedge (where $V$ is half a spatial hyperplane). $S(\rho_V)$ would be given naively by the entropy in a thermal gas at the Unruh temperature $T=(2\pi\, z)^{-1}$, where $z$ is the distance to the boundary plane. For massless particles the entropy density is then proportional to $z^{-3}$ and the entropy diverges proportionally to the integral of this quantity in half space, giving $S\sim \epsilon^{-2}{\cal A} $, where ${\cal A}$ is the area on the boundary surface, and $\epsilon$ is a short distance cutoff. 

This type of divergences is a general feature of the localized entropy. This is because the localization gives place to pair creation in the boundary region.  
For a general set $V$ we have an expansion of form \cite{ch2,fur}
 \begin{equation}
 S(\rho_V)=g_{2}[\partial V]\,\epsilon^{-2}+ g_1[\partial V]\,\epsilon^{-1} + g_0[\partial V]\,\log (\epsilon)+ S^F(\rho_V)\,,   \label{div}
 \end{equation}
 where $S^F(\rho_V)$ is a finite part, and the $g_i$ are local and extensive functions on the boundary $\partial V$, which are homogeneous of degree $i$.  These divergent terms are ambiguous since they depend on the cutoff procedure (excepting the coefficient of the logarithmic term \cite{ch2}

To avoid the divergences one can put the system in a box to enforce boundary conditions and eliminate the entanglement with the exterior region. However, the box itself would be made of some material, and the consideration of the relevant fields for it would take us back to the same problem. Enforcing artificial boundary conditions cannot take away the ambiguities.

However, the entropy one actually wants to measure is the one in the system under consideration on the region $V$ and not the one due to the cloud of the vacuum fluctuations, which is present for any state. This is the crucial point made in \cite{maro}. Thus one should subtract from the system entropy the entropy in the vacuum state $\rho_V^0=\textrm{tr}_{-V}\,\left| 0 \right> \left< 0 \right|$, corresponding to the same region $V$. It is important to note here that since the divergences have an ultraviolet origin the divergent terms in (\ref{div}) are purely geometric and should not depend on the state $\rho$ (at least if this later has a physically acceptable ultraviolet behavior, for example if it has finite energy). Thus, while both entropies contain divergences, the difference between the local entropies for two states is well defined, independent of the regularization. Our candidate for the left hand side of (\ref{uno}) is then 
\begin{equation}
S_V=S(\rho_V)-S(\rho^0_V)\,.
\end{equation}

When the bulk of the system is far from the boundaries of $V$ the entropy $S_V$ should approach the global state von Neumann entropy $S(\rho)$. We think a formal proof of this could be attempted, for example considering the limit of sets with increasing sizes, or, for the Rindler wedge, the limit of large translations of the state $\rho$.
A simple example was given in \cite{maro}, where the entropy of a mixed state between different species of a single scalar particle in half space is computed. The result was there given in terms of a series. This is reviewed in the appendix, where we have also provided an analytic expression for the series sum.

\section{A localized energy}
Now we turn to the right hand side of (\ref{uno}). Again we find difficulties in trying to define the quantities involved. The first one is that, strictly speaking, the energy of a localized state like $\rho_V$ cannot be computed, since the Hamiltonian operator is defined in the whole Hilbert space, while the local density matrices act on the operator algebra generated by the observables localized in $V$. 
For this reason we need to define a local form of energy which has sense in the restricted space. 

Also, there is no general consensus on the correct value of $\lambda$ ($2\pi$ from the original derivation \cite{bekbb}, somewhat larger in \cite{bekdowe}) or the meaning of $R$ (circumscribing radius of the system in \cite{bekbb}, the shorter dimension of the system in \cite{bouchica}).

What we propose is in fact a candidate for the product $\lambda \,E\,R$ in (\ref{uno}). In this way we eliminate the ambiguities in the different quantities by a good definition of their product\footnote{A similar step was carried out in \cite{armonic} by using the harmonic resolution instead of the product $E\,R$. However, in the present case we do not want to abandon the localization of the states, which we consider an essential ingredient in the derivation of the bound from the GSL.}. 

In order to proceed we take the local vacuum density matrix (we still think in terms of a cutoff theory) and write it as
\begin{equation}
\rho^0_V=\frac{e^{-\, K}}{\textrm{tr}\,e^{- \, K}} \,.\label{form}
\end{equation} 
The hermitian and positive definite matrix $\rho_V^0$ can always be written in this way since it has no zero eigenvalues (except when $V$ is the whole space, where it becomes the vacuum state projector). The Hermitian and dimensionless operator $K$ can be called the local Hamiltonian for $V$. This operator can also be defined directly in the continuum theory, where it is called the modular Hamiltonian \cite{haag}. Note that the operator $e^{-i K s}\sim (\rho_V^0)^{i s}$ is unitary, and acts on the algebra of fields in $V$ as a kind of local dynamics with time $s$.  

  When $V$ is half a spatial plane (its causal development is the Rindler wedge), the form of $K$ is known exactly. In this case $\rho_V^0$ can be written in the form (\ref{form}) with $K$ proportional to the boost operator which keeps the wedge invariant \cite{unruh}. Explicitly, 
\begin{equation}
K=2\pi \int dx dy \,\int_0^\infty dz \,\,  z\,\,{\cal H}(x,y,z,0) \,,\label{boo}
\end{equation}  
where $x$, $y$ and $z$ are the spatial coordinates, $V$ is the set $z\ge 0$, and ${\cal H}(x,y,z,t)$ is the Hamiltonian density operator. 
 The local dynamics is then given by the trajectories of constant acceleration observers
 \begin{eqnarray}
 t^\prime&=&\cosh (2\pi s)\,t +  \sinh (2\pi s) \, z\,,\nonumber \\
 z^\prime&=&\sinh (2\pi s) \, t+ \cosh (2\pi s)\,z \,. 
 \end{eqnarray}
  Remarkably, this holds for any QFT. 
  
  Our proposal for the product $\lambda E R$ suggest itself from the formula (\ref{boo}) and is given by the expectation value of the operator $K$. 
  When most of a system which has a size $L$ is far from the boundary, $R\gtrsim L$, this gives approximately $K\sim 2\pi E\,R$.   
  
   As a different example for $K$ consider the case of conformal field theories (in any dimensions), and when the set $V$ is a sphere \cite{sphere}. Then $K$ is proportional to the generator of the conformal transformations which keep the sphere fixed. In polar coordinates it writes 
\begin{equation}
K=2\pi \int_0^R dr \int r^2\, d\Omega \,  \frac{(R-r)(R+r)}{2R}\,\,{\cal H}(r,\theta,\phi,0) \,.
\end{equation} 
 The local dynamics is shown in the figure 1 and is given by    
\begin{equation}
X_\pm^\prime=R\,\frac{\left( (R+X_\pm)-e^{2\pi s} (R-X_\pm)\right)}{\left( (R+X_\pm)+e^{2\pi s} (R-X_\pm)\right)} \,, \label{fl}
\end{equation}
where $X_+=t+r$ and $X_-=r-t$ are null coordinates, and here $R$ means the radius of the sphere.  

\begin{figure}[t]
\centering
\leavevmode
\epsfysize=5cm
\epsfbox{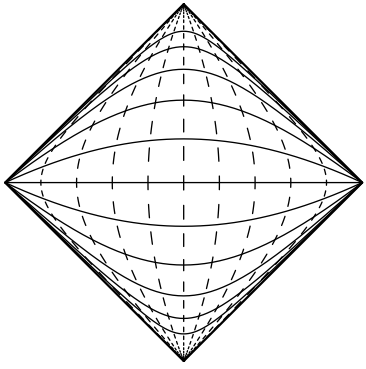}
\bigskip
\caption{The horizontal interval represents a spatial sphere (only one spatial dimension shown), while the diamond shaped set is its domain of dependence. The vertical dashed curves are the trajectories of the point transformations induced by the local dynamics in a conformal field theory. The horizontal solid curves are the translation of the sphere for different internal times $s$ (from bottom to top, the curves correspond to integer $s$ from $-5$ to $5$).}
\end{figure}

Near the boundaries this flow approaches the one for the half space, and this feature is expected to be generic for any QFT and any set with smooth boundaries. Far from the boundaries the modular flow is not expected to be universal. For example, in the center of the sphere $r=0$, at a distance $R$ from the boundary, the velocity of the flow (\ref{fl}) is half the one corresponding to a distance $R$ from the boundary in the Rindler wedge. That is, one has $K\sim \pi E R$ for a state well localized near the center of the sphere.   

For massive or interacting theories, or sets of more complex shape, the modular Hamiltonian is not expected to be local, and the local dynamics does not act as a simple space-time point transformation on the fields. For example, for a free massive Dirac fermion we have \cite{massive}
\begin{equation}
K=\int_V dx dy \,\Psi^\dagger(x)\log\left(C_V^{-1}-1\right)(x,y)\,\Psi(y)\,,
\end{equation}
where
\begin{equation}
C_V(x,y)=\left<0|\Psi^\dagger (x) \Psi (y)|0 \right>
\end{equation}
 is the field correlator inside the set $V$. This Hamiltonian is still quadratic, but it is in general non-local with respect to the fermion fields. 

A local Hamiltonian can also be defined for other states and space-times, in a way analogous to (\ref{form}). One important case where it is possible to compute it exactly is when the state on $V$ is thermal with respect to a given notion of time. In this case the reduced density matrix is of the form 
\begin{equation}
\rho_V=\frac{e^{-\beta \, E}}{\textrm{tr}\,e^{-\beta \, E}} \,,\label{trece}
\end{equation}
with $E$ the time translation operator, and $\beta$ the inverse temperature. Then, we have comparing with (\ref{form})
\begin{equation}
K=\beta \,E\,.
\end{equation}
This is the case for a black hole in the Hartle-Hawking state. For a Schwarzschild black hole we have in this case  
$K= 8\pi G M \,E$, where $E$ is the energy operator as measured by asymptotic observers, and $M$ the black hole mass. More explicitly, in this case we have 
\begin{equation}
K=8\pi GM \,\int_V dv \, \left(\frac{r-2M}{r}\right)^{\frac{1}{2}}\, {\cal H}(r,\theta, \phi,0)\,, \label{kbh}
\end{equation}
where ${\cal H}=T_0^0$ is the Hamiltonian density in Schwarzschild coordinates, and the integration is over the exterior of the black hole region. 
As in the Rindler case, this is local and has a universal expression which is in principle valid for any theory. As it is well known, this operator approaches (\ref{boo}), proportional to the boost operator in the Rindler space, very near the horizon. For this we have to pass to distance coordinates, $dz=(1-2MG/r)^{-(1/2)}dr$, and take the near horizon limit, $(r-2MG)\ll 2MG$. In the opposite limit, $r\gg 2MG$, $K$ is proportional to the usual Minkowski space time translations operator.   

Returning to flat space-time, and to the vacuum state induced local Hamiltonian $K$,  there is one further problem with taking its expectation value on the localized states which we have not mentioned yet. To be concrete, consider the Rindler wedge. The vacuum state is thermal with respect to the boost operator with temperature $(2\pi)^{-1}$. Thus, heuristically, for a massless field we can think in a thermal state for the usual time translations with temperature varying as $T\sim 1/z$, and then "boost energy" density going as $z^{-3}$. This diverges when integrated near the boundary. Another aspect of the same problem follows from the expression (\ref{form}) for the density matrix in terms of $K$. The replacement $K\rightarrow K+ \textrm{constant}$ does not change $\rho_V^0$. This makes $K$ (and its expectation values) ambiguous by redefinitions by a constant term.  

 Both of these problems can be cured by subtracting the corresponding vacuum expectation value of $K$ in order to eliminate the divergent contribution coming from the vacuum fluctuations. The proposal for the right hand side then reads
\begin{equation}
K_V=\textrm{tr} ( K \rho_V )-\textrm{tr} ( K \rho^0_V )\,.
\end{equation} 
As happens for the entropy, this subtraction should lead again to a well defined cutoff independent quantity.

\section{Relative entropy and the Bekenstein bound} 

Our version of the Bekenstein bound then reads $S_V\le K_V$, which is
\begin{equation}
S(\rho_V)-S(\rho^0_V)\le \textrm{tr} ( K \rho_V )-\textrm{tr} ( K \rho^0_V )\label{esta}\,.
\end{equation}
According to (\ref{form}) we have $K=-\log \rho^0_V-\log(\textrm{tr}\,e^{-K})$. Using this and $\textrm{tr} \rho_V=\textrm{tr} \rho^0_V=1$, we can write (\ref{esta}) simply as
\begin{equation}
\textrm{tr}(\rho_V \log \rho_V)-\textrm{tr}(\rho_V \log \rho^0_V)\ge 0\,.
\end{equation}
This is just the statement  of the positivity of the relative entropy $S(\rho_V |\rho_V^0)$ between the local density matrices corresponding to the system's state and the vacuum, both reduced to $V$. 

The relative entropy between two density matrices $\rho_1$ and $\rho_2$ is defined as
\begin{equation}
S(\rho_1 \vert \rho_2)=\textrm{tr}(\rho_1 \log \rho_1)- \textrm{tr} (\rho_1 \log \rho_2)\,,\label{rela}
\end{equation}
and has the meaning of the statistical distance between the pair of states\footnote{In quantum field theory  $S(\rho_V \vert \rho_V^0)$  is perfectly well defined, independently of any cutoff procedure, even if the existence of the density matrices requires regularization. This mathematical definition is achieved by the Araki formula for the relative entropy for a pair of states in the local operator algebras \cite{araki}. According to the discussion in the previous sections the quantities $S_V$ and $K_V$ should also have a direct definition in the continuous QFT (perhaps it is even posible to define them for an arbitrary von Neumann algebra). However, to the author's knowledge these mathematical definitions have not been worked out in the literature. This interesting point deserves further study.} \cite{relative}. It is always  positive  
\begin{equation}
S(\rho_1 \vert \rho_2)\ge 0\,,
\end{equation}
and is zero only when $ \rho_1 =\rho_2$. This gives a proof of the present  formulation of the bound. 
As an example, for an object far enough from the boundaries in a half space the relation (\ref{esta}) approaches (\ref{uno}) with $\lambda=2\pi$.
In this sense (\ref{esta}) captures what is underlying the validity of the bound in several examples studied in the literature.

The relative entropy has also a well known application to physics in relation with the free energy. If the state $\rho_2$ is thermal with respect to the Hamiltonian $E$ as in (\ref{trece}),
its relative entropy with the state $\rho_1$ is 
\begin{equation}
S(\rho_1 \vert \rho_2)=\beta (F(\rho_1)-F(\rho_2))\,,\label{free}
\end{equation}
where the free energy $F(\rho)$ is given by
\begin{equation}
F(\rho)=\textrm{tr} (\rho E )- T S(\rho) \,.
\end{equation}
Therefore, the positivity of the relative entropy is equivalent to the fact that the free energy of the thermal equilibrium state is the minimum one.

One interesting point is how this formulation of the bound manages to avoid the species problem. The reason was first discovered in \cite{maro}. The species problem in the Bekenstein bound arises since if the number of independent particle species $M$ is increased we can make the entropy of $\rho_V$ to increase keeping its energy fixed. One example is to take $\rho_V=M^{-1}\sum \rho_V^i$, where $\rho_V^i$ represents a state containing only particles of the $i$ type. The $\rho_V^i$ for different $i$ can be taken to be the same state with only the particles species interchanged. The entropy $S(\rho_V)$ will then increase with $M$. However, the reduced vacuum state $\rho_V^0=\otimes \rho_V^{0i}$ also becomes more entropic, and even if $S(\rho_V)$ and $S(\rho_V^0)$ both increase with $M$, their difference saturates for large enough values of the species number (a particular example of a single particle was studied in \cite{maro} and is reviewed in the appendix). 
Moreover, 
the relative difference between the states, as measured by their relative entropy, diminishes with $M$, because it becomes more difficult to separate the state from the thermal atmosphere generated by the localization. To be more explicit, using the property of joint convexity 
\begin{equation}
S\left(\sum \lambda_i\rho_1^i\,\vert\, \sum \lambda_i \rho_2^i\right)\le \sum \lambda_i S(\rho_1^i|\rho_2^i)\,,
\end{equation}
we have for the particular states considered above 
\begin{equation}
 0 \le S\left(M^{-1}\sum \rho_V^i\vert \rho_V^0\right)\le  S( \rho_V^i \vert \rho_V^{0i})\,.
\end{equation}
 The interpretation of these inequalities is that the relative entropy measures the degree of distinguishability between the two states \cite{relative}, and, in contrast to the entropy, decreases under mixing, making the bound more strict. However, the bound never get violated since when the states are totally undistinguishable the relative entropy is zero. 
 
Another question is which region $V$ makes the bound more severe for a given state. We have shown above that the sphere can make a better bound than half the space. In general we have that the relative entropy is monotonously increasing with size
\begin{equation}
S\left(\rho_V\vert \rho_V^0\right)\le S\left(\rho_W\vert \rho_W^{0}\right)\,, \hspace{2cm} V\subseteq W\,. \label{mono}
\end{equation}
Therefore its smallest values hold for small sets, where the state $\rho_V$ resembles the vacuum and $S\left(\rho_V\vert \rho_V^0\right)$ tends to zero. For large sets which include most of the state $\rho$ it would increase without upper bound with the typical distance $R$ between the set boundary and the system. 
Perhaps a more interesting and more difficult way to state this question is for which set the ratio $S_V/K_V\le 1$ achieves a maximum for a given state. It is possible that in this case the set must adjust itself to some "form of the state", enclosing most of the entropy at the minimum distance.

\section{Discussion: the Bekenstein bound and the GSL}

The Bekenstein bound follows from the generalized second law by the thought experiment involving classical physics described in the introduction. However, there are, at the semiclassical level and in the quasistacionary regime, at least two different proofs of the GSL. These do not make use of a new principle like the Bekenstein bound, suggesting that properly including the quantum corrections the bound must hold necessarily. The purpose of this section is to make contact with these proofs. 

The Sorkin's proof for the semiclassical quasistatic case \cite{sork2} relies only on the region of the space-time external to the black hole, which has autonomous evolution (this proof must not be confused with the more general proof-scheme of the GSL in quantum gravity by the same author given in \cite{sork1}). Let $\rho_1$ and $\rho_2$
 be the two external states corresponding to the two different times $t_2\ge t_1 $, characterized by two Cauchy surfaces $\Sigma_1$ and $\Sigma_2$. Sorkin also assumes the existence of the Hartle-Hawking thermal equilibrium state for the exterior region, $\rho_{\textrm{HH}}\sim e^{-\frac{E}{T}}$, with $T$ the black hole temperature and $E$ the time translations symmetry operator for the static geometry. Here $E/T$ is also the local Hamiltonian for the static exterior geometry in thermal equilibrium. Under the quasistatic approximation we can make the assumption that the time translation invariant state $\rho_{\textrm{HH}}$ is the same on $\Sigma_1$ and $\Sigma_2$. If the time evolution for the external states is completely positive-trace preserving (CPT) \cite{relative} we have 
\begin{equation}
S(\rho_1|\rho_{\textrm{HH}})\ge S(\rho_2|\rho_{\textrm{HH}})\,.\label{qqq}
\end{equation}
In fact, the CPT property in this case holds, and coincides with the monotonicity for the operator algebras (\ref{mono}), taking into account that the state $\rho_2$ is obtained from $\rho_1$ by unitary evolution from $t_1$ to $t_2$ plus trace over the degrees of freedom which on $\Sigma_2$ lie inside the horizon. This inequality is equivalent to 
\begin{equation}
S(\rho_2)-S(\rho_1)-\frac{(\left<E\right>_2-\left<E\right>_1)}{T} \ge 0\,,\label{ppp}
\end{equation}
and using the quasistatic relation $\Delta S_{\textrm{BH}}=(\left<E\right>_1-\left<E\right>_2)/T$, gives the GSL. 

Now we turn to the experiment of letting an object be swallowed by the black hole.  The least constraining inequality from (\ref{qqq}) is just $S(\rho_1 |\rho_{\textrm{HH}})\ge 0$, which happens for $\rho_2=\rho_{\textrm{HH}}$. This gives our proposal (\ref{esta}) for the Bekenstein bound, but here the role of the vacuum is played by the Hartle-Hawking state. More explicitly, in this case (\ref{ppp}) gives 
\begin{equation}
S(\rho_1)-S(\rho_{\textrm{HH}})\le \textrm{tr}\left(\frac{E}{T}\, \rho_1\right)-\textrm{tr}\left(\frac{E}{T}\, \rho_{\textrm{HH}}\right)\,.\label{tb}
\end{equation}
Both members of this inequality should be finite. As mentioned, the combination $E/T$ is just the local Hamiltonian (\ref{kbh}) for the external black hole geometry, and for objects near the horizon it approaches the corresponding one for the Rindler wedge. Thus, in the large black hole-flat space limit, where the space-time curvature can be neglected, we have the inequality (\ref{esta}), where $K$ is the local Hamiltonian of (\ref{boo}) corresponding to the Rindler wedge. Therefore, the conclusion is that a form of the Bekenstein bound is indeed required by the GSL, but it is such that it does not introduce new physical constraints to QFT.   
 
The relation (\ref{tb}) is a particular case of a more general one, where the Hartle-Hawking state is replaced by the thermal equilibrium state at temperature $T$ ((see (\ref{free})). The interpretation of the inequality in this case is just that any object may be made to come to thermal equilibrium with a thermal bath without violating the ordinary second law. 

However, in the thought experiment leading to the Bekenstein bound, the localization of the states introduces a very important difference: crossing the horizon means in this language to be absorbed by the thermal reservoir, and not just to be put into thermal contact with it. This may be interpreted as a sudden change in the Hamiltonian, rather than an interchange of energy, and in this case the result of the experiment cannot be explained thermodynamically. As the metric is the same for any type of matter, it may allow any object to be swallowed by the black hole (according to several works \cite{crit, marosork} this is a weak point of the Bekenstein deduction of the bound from the second law, since under certain conditions this could be impeded to occur). But at the same time, this same universal character of the metric gives place to a thermal atmosphere for any type of matter, which is already present alongside with the object initial state, and will also be present in the final state. The thermodynamic interpretation is then recovered by thinking in the autonomous exterior region alone, where an object and the thermal atmosphere together are seen to approach the state of thermal equilibrium. 
The Bekenstein bound then resolves in that for very entropic objects a very entropic thermal atmosphere is expected, and one has to take into account its entropy for the external region. The importance of thermal atmosphere then relies not on its possible mechanical effects on the object, but rather on the information available to the observer, which in some situations may not easily distinguish the object from the thermal cloud \cite{maro}. Note also that the thought experiment described above can be repeated word by word for any set $V$ in flat space, where the exterior region is taken as the domain of dependence of $V$ and $\rho_{\textrm{HH}}$ is replaced by $\rho_V^0$.

Another proof of the GSL has been given by Frolov and Page \cite{page}. They consider the eternal black hole geometry for the quasistacionary process. Without entering into the details, this proof, apparently different in nature to the one by Sorkin, also boils down to the positivity of a particular relative entropy
\begin{equation}
S(\rho_1|\rho_\textrm{HH})\ge 0\,,
\end{equation} 
where here $\rho_1$ is the part of the incoming state which goes down to the future horizon, and $\rho_\textrm{HH}$ is the outgoing thermal state on the past horizon. The future and past horizon Hilbert spaces are mapped to each other by time reflection symmetry. Except for considering the horizon states instead of the exterior states, the conclusion is the same as above. 

 A word of caution about the range of validity of these proofs (see also \cite{bekdowe}).  Strictly speaking, the relative entropy for two states in a given region can only be defined in a fixed geometry, and becomes ill defined once the backreaction is included. This is because two different states correspond in general to two different metrics, making the quantities in (\ref{rela}) uncertain (or UV divergent) \cite{mutual}. Then, in both of these proofs the static limit is crucial, what means a flat space limit for the Bekenstein bound. For a more general situation, but still in the semiclassical case, it is possible that we have to resort to express the quantities involved in terms of the mutual information \cite{mutual}.

\section{Appendix: a scalar particle in the Rindler wedge}
In this appendix we review a calculation of the subtracted entropy $S_V$ by Marolf et al \cite{maro}, which is very instructive regarding the solution of the species problem for the Bekenstein bound.  We also give an analytic expression for their final result for $S_V$, and express it in terms the relative entropy, demonstrating the validity of the approximations used in that work.

 Consider a state consisting in a superposition with equal weights of a single particle state corresponding to $M$ different free scalar fields. We compute the relative entropy between this mixed state and the vacuum, both reduced to $V$, which is taken to be half a spatial hyperplane. 

To begin with, we recall that in this case the Hilbert space can be decomposed into a tensor product of the states in the left half space and on the right one, ${\cal H}={\cal H}_L\otimes {\cal H}_R$. Accordingly, the Minkowski vacuum state for each of the $M$ fields decomposes as \cite{unruh}
\begin{equation}
\left| 0\right>=\Pi_{\omega,k} (1-e^{-\omega})^{\frac{1}{2}} \sum_{N=0}^\infty e^{-\frac{\omega N}{2}}\left|N,N \right>\,.
\end{equation}
Here some discretization of the wave packets is assumed. The dimensionless variable $\omega$ is $2\pi$ times the eigenvalue of the boost operator, while $k$ represents the additional quantum numbers which define the wave packets. 

Following \cite{maro} we make the simplifying assumption that the single particle mode which we are considering, in addition to have positive Minkowski energy, is well localized inside the right half space.  This localization forces one to take $\omega\gg 1$. Then we can express the Minkowski annihilation operator for the mode to an exponentially small error as 
\begin{equation}
a=\frac{(a_R^w-e^{-\frac{\omega}{2}} a_L^{w\dagger})}{\sqrt{1-e^{-\omega}}}\,.
\end{equation} 
In this equation all the wave packet creation operators are normalized in standard way, $[a,a^\dagger]=1$. This mode then has boost energy highly peaked around $\omega/(2\pi)$.
This expression neglects contributions proportional to $a_R^\dagger$ and $a_L$ which are exponentially small in $\omega$. We have retained the term proportional to $a_L^\dagger$ in order to have $a\left|0\right>=0$. 

 The problem then decomposes  into the different frequencies $w$. A finite entropy for the reduced states require the presence of an infrared and an ultraviolet cutoff in order to make finite the sum over the modes. However, for the quantities we are interested in, which involve differences between the state and the vacuum, the only relevant contribution comes from the mode which is excited in the one particle state with respect to the vacuum. Then, for ease of notation, in what follows we omit all other modes from consideration. With this convention the single particle state is 
 \begin{equation}
 a^\dagger \left|0\right>=(1-e^{-\omega})\sum_{N=0}^\infty \sqrt{N+1}e^{-\frac{\omega N}{2}}\left|N,N+1 \right>\,.
 \end{equation}
Now consider $M$ independent fields. The vacuum state is $\left|0\right>=\otimes_i \left|0\right>_i$, and we take for the particle the mixed state $M^{-1/2}\sum_i a_i^{\dagger} \left|0\right>$.
From here, tracing over the left Hilbert space, we obtain the reduced density matrices. Adopting a vector notation $\vec{N}=(N_1,...,N_M)$ and calling ${\cal N}=\sum_{k=1}^M N_k$ we have 
\begin{eqnarray}
\rho_V^0&=&  (1-e^{-\omega})^M \sum_{\vec{N}=0}^{\infty}   e^{-\omega {\cal N}}\left|N_1,...,N_M \right>  \left<N_1,...,N_M \right|  \label{teta}\,,\\
\rho_V&=&\frac{1}{M}e^{\omega}(1-e^{-\omega})^{M+1}\sum_{\vec{N}=0}^{\infty}  {\cal N} e^{-\omega {\cal N}}\left|N_1,...,N_M \right>  \left<N_1,...,N_M \right|\,. \label{tota}
\end{eqnarray}
Both of these are diagonal. 
The entropy of the vacuum corresponding to the mode in consideration is
\begin{equation}
S(\rho_V^0)=M\left( \frac{w}{e^\omega-1}-\log\left(1-e^{-\omega}\right)\right)\,,
\end{equation}
which is the one of a thermal ensemble of M independent oscillators. 
 It also follows that 
\begin{equation}
S(\rho_V)=-\sum_{\vec{N}=\vec{0}}^\infty \frac{1}{M}e^{\omega}(1-e^{-\omega})^{M+1} {\cal N} e^{-\omega {\cal N}} \log\left(\frac{1}{M}e^{\omega}(1-e^{-\omega})^{M+1} {\cal N} e^{-\omega {\cal N}} \right)\,. \label{tt}
\end{equation}
This is the expression found in \cite{maro}. It can be easily evaluated excepting for the term proportional to $\sum_{\vec{N}=\vec{0}}^\infty      {\cal N} e^{-\omega {\cal N}}    \log\left({\cal N}\right)$. In order to obtain an analytic expression for this sum we use the integral representation for the logarithm
\begin{equation}
\sum_{\vec{N}>\vec{0}}^\infty  e^{-\omega {\cal N}}    \log\left({\cal N}\right)=\int^\infty_{1} da \sum_{\vec{N}>\vec{0}}^\infty  e^{-\omega {\cal N}}\left(\frac{1}{a}-\frac{1}{{\cal N}-1+a}\right)\,.\label{hh} 
\end{equation}
Then, defining 
\begin{equation}
g(\omega,a)=\sum_{\vec{N}>\vec{0}}^\infty  \frac{e^{-\omega {\cal N}}}{{\cal N}-1+a}\,,
\end{equation}
and deriving with respect to $\omega$ we have the differential equation
\begin{equation}
\frac{d g(\omega,a)}{d\omega}-(a-1)g(\omega,a)+\frac{1}{(1-e^{-\omega})^M}-1=0\,.
\end{equation}
Taking into account the boundary condition $\lim_{\omega\rightarrow\infty}g(\omega,a)\rightarrow 0$, the solution is
\begin{equation}
g(\omega,a)=-e^{(a-1)\omega}\int^\infty_\omega \, dy\, e^{(1-a)y}(1-(1-e^{-y})^{-M})\,.
\end{equation} 
Using this result in (\ref{tt}) and (\ref{hh}) we obtain
\begin{eqnarray}
S_V&=&S(\rho_V)-S(\rho_V^0)= \nonumber \\
&&\log(M)-\log(e^\omega-1)+\frac{\omega}{1-e^{-\omega}}+\int_0^\infty du \frac{e^{-u}}{u}\left(\left(\frac{1-e^{-\omega}}{1-e^{-\omega-u}}\right)^{M+1}-1\right)\,.\label{aquella}
\end{eqnarray}

If we consider the number of species $M$ fixed and take the limit of a particle far from boundary, such that $M e^{-w}\ll 1$, we have
\begin{equation}
S_V\sim \log (M) + {\cal O} (M e^{-w}) \,.
\end{equation}
In this limit the finite local entropy $S_V$ coincides with the global state entropy $\log (M)$.  From (\ref{aquella}) it follows that the entropy increases monotonically with $M$. If the behavior for large $M$ were still logarithmical, at some point the Bekenstein bound would get violated. However, from (\ref{aquella}) we see that the entropy saturates in the limit of large species number and fixed $w$,  $M e^{-w}\gg 1$,
\begin{equation}
S_V\sim \frac{w}{1-e^{-w}}+{\cal O} (M^{-1}e^{w})\,.\label{par}
\end{equation}

A similar calculation gives for the relative entropy between (\ref{teta}) and (\ref{tota})
\begin{equation}
S(\rho_V|\rho_V^0)=\frac{w}{1-e^{-w}}-(S(\rho_V)-S(\rho_V^0))\,.
\end{equation}
The positivity of the relative entropy then also shows that (\ref{par}) is an upper bound on the local entropy $S_V$ no matter the number of species. This  ensures the Bekenstein bound is not violated.

\end{document}